# Tetrahedral Units:
## forDodecahedral Super-Structures

Y. Ortiz [a,b], D. J. Klein[a], & J. F. Liebman[c]

[a] MARS, Texas A&M University at Galveston, Galveston,
200 Seawolf Pkwy, Galveston, TX 77554
[b] Instituto de CienciasFísicas, Universidad NacionalAutónoma de México,
Avenida Universidad s/n, 62210 Cuernavaca, México;
[c] Dept of Chemistry & Biochemistry, University of Maryland, Baltimore County,
1000 Hilltop Circle, Baltimore, MD 21250

**Abstract**

Different novel organic-chemical possibilities for tetrahedral building units are considered, with attention to their utility in constructing different super-structures. As a representative construction we consider the useof sets of 20 such identical tetrahedral units to form a super-dodecahedron.

**Introduction**

Recently Minyaev*et al* [1] proposed a dodecahedral-symmetry molecule comprised of tetrahedral submolecular building blocks placed at the corners of a large dodecahedron with the tetrahedral units interconnected via covalent bonds. The tetrahedral units in isolation are of the form $TH_4$ with the 4 H atoms bonded covalently to the central tetrahedral core T. Then taking 20 of these tetrahedral building blocks, one deletes 3 H atoms from each, and reconnects the consequent dangling bonds of each resultant TH to 3 of these equivalent neighboring TH units, to obtain $(TH)_{20}$– as indicated in **Fig**.1.The prototype for this construction may be viewed to be Paquette*et al*'s[2]dodecahedrane, $(CH)_{20}$, with T being just a single C atom. The tetrahedral units considered by Minyaev*et al*[1] aretetrahedra where T is taken as a (hydrogen-deleted) cage: $B_4$, or $C_4$, or $Al_4$. In fact, there are many more (hydrogenated) tetrahedral cages $TH_4$ possible which may be utilized in such a construction of a "super-dodecahedron". One such is the well-known adamantane, which indeed has in fact earlier been proposed [3] as a tetrahedral construction unit for making a super-dodecahedron, $(TH)_{20}$, or even a super-truncated-icosahedron,$(TH)_{60}$– that is, $TH_4$isadamantane, $(CH)_4(CH_2)_6$, with the 4 exceptional H atoms being those attached to the tertiary carbons, so that T is $C_4(CH_2)_6$. This adamantane tetrahedral unit has also been described[4] as a possibility to form a tetrahedral"hyper-adamantane", as well as a host of other nominally "unstressed"structures– whence one could consider thishyper-adamantane as a new tetrahedral building block. But there are yetmany furtherpossibilities for tetrahedral building blocks. Recently also Huang *et al*[5] emphasize the use of tetrahedral building blocks (especially based on silessquioxanes) to build up a variety of nano-structures. Indeed the general idea of using a convenient type of building-block unit (a molecular tinker-toy unit [6]) may be viewed as useful idea for nano-technology.

Here we indicate other possible tetrahedral building blocks, of organic-chemical form satisfying standard bonding schemes, for stability. We indicate first a collection of adamantane-related such building blocks, then aclass [7] of tetrahedral fulleroid building blocks, then a further class of adamanto-capped fullerenes. Each of these novel classes manifest infinite sequences of ever larger tetrahedral units, and it is noted how suitable units of higher than tetrahedral symmetry may be used as tetrahedral building blocks. As an example target larger nano-structure we consider the building of a super-dodecahedron from such tetrahedral units.

Yet further the feature that in placing tetrahedral building blocks into the super-dodecahedral arrangement, there is some bond-angle strain, since the $108°$bond angles of a dodecahedron do not precisely match the $\approx 109.5^d$ tetrahedral angles. Here what we mean by strain & stress should be distinguished from another popular nomenclatural choice, advocated by Schleyer*et al* [8] and adopted by most organic chemists. Basically they make a choice of a type of group function expansion, to reveal higher order non-neighbor corrections, which are chosen to be designated as measures of what they term "stress & strain" – their result being that though the bond angles and bond lengths are essentially ideal in adamantane, they still compute a non-negligible energy cost, which they call a "strain energy". But there is little relation to the standard definitions from elementary physics & engineering where strain is to measure some geometric deviation, with a stress-energy cost (quadratic in the strain) – indeed with Schleyer *et al*'s nomenclature one is led to the idea of "strain energy" in the absence of (geometric) strain. Moreover, it is to be emphasized that there is at least one alternative [9] to Schleyer *et al*'s choice of group-function expansion to lead to quite different results, e.g. adamantane was asserted to be strainfree. Thus here we prefer to refer to Schleyer *et al*'s computed energies as "tension energies", and further refer to "strain" in a geometric context, measuring geometric deviations – angles in the currently noted case – while stress refers to an energetic consequence, anticipated to be quadratic in the strain. Granted this, a way to ameliorate the modest consequent (energetic) stress in our super-dodecahedranes (including the classical dodechedrane $C_{20}H_{20}$) is indicated. Some exemplary computations for such super-dodecahedra are made, noting a signature of the stress in the vibrational-frequency spectrum.

**Adamantane-based Tetrahedral Units**

Adamantane $C_{10}H_{16}$ itself has the structure of **Fig. 2**, with the 4 tetrahedrally located H atoms in bold-face. **Fig. 3** shows super-adamantane, as previously described, to result from taking 10 adamantanes, placing them at corners of an enlarged adamantane skeleton, then appropriately deleting H atoms, and interconnecting. There are two intermediate possibilities, where either just the 4 tertiary C-atoms are replaced by adamantanes, or the 6 secondary C-atoms are replaced by adamantanes – as in **Fig.4**.

There are further possibilities, using one of the structures of **Fig.2**&**3** as tetrahedra to be placed in a larger adamantane structure. That is, if we have 2 tetrahedra $TH_4$ & $T'H_4$, then one can form a new super-adamantyltetrahedron $(TH)_4(T'H_2)_6$. In more systematic detail for our adamantane-based constructions, we can define a replacement operation $a$ which applied to a pair of tetrahedral structures T & T' places 6 copies of T & 4 copies of T' respectively in the 6 secondary & 4 tertiary positions of an enlarged adamantane structure – the result being denoted $a(T,T')$. Thus denoting the single carbon atom choice by g, we then have:

adamantane = $a(g,g) \equiv a$

hyper-adamantane (Fig. 2) = $a(a(g,g),a(g,g)) \equiv a_{aa}$

Fig 3a = $a(g,a) \equiv a_{ga}$

Fig 3b = $a(a,g) \equiv a_{ag}$

hyper-hyper-adamantane (Ref. 4) = $a(a_{aa},a_{aa})$

This last one has 3 *levels* of the $a$-replacement operation, though there are several other species with 3 levels of this operation –in fact 20 more. The consequent such species down to 3 levels are indicated in Table **I**, where $\alpha_n$ denotes the number of $a$-transformations at level $n$ of a considered structure. Clearly the number of such iterated adamantyl structures thru to a level $n$ increases quite rapidly with $n$ – as indeed can be expressed[a] in a quantitative fashion – to reveal a super-exponential number of structures having up to $n$ levels.

**Fulleroid Tetrahedral Units**

These tetrahedral species $TH_4$ of a general hydro-fulleroid class are to be constructed by a general means proceeding thru a sequence of simple constructions. We start from the graphene (or honeycomb) net, and begin the construction from the center $a$ of one of the hexagonal rings. For the $(h,k)$ th member of the class, now take a straight walk center-to-center thru $h$ rings, and then turn 60° taking a straight walk thru $k$ rings, to end up in a second ring center $b$. Then from $b$ a similar sort of walk of $h$ steps (at 120° to the direction of the first $h$-step walk from $a$) followed by $k$ steps rotated 60° more. This ends up at a third hexagon center $c$, which is such that if one now takes an $h$-step walk from $c$ in a direction at 120° to that from $b$, followed by a further 60°-rotated $k$-step walk, then one ends up at $a$. This construction is indicated in **Fig.6**, for $(h,k) = (2,1)$. Granted the 3 centers $a,b,c$ on a regular honeycomb net, one now imagines cutting out an equilateral triangle from the net, with vertex-locations at these 3 centers. Next taking 4 of these equilateral triangles with their fragments of the net, they are imagined to be pasted to the faces of a tetrahedron, thereby giving a "fulleroid" with triples of equilateral corners of the $(h,k)$ triangles meeting at tetrahedron corners, each of which is surrounded by a triangular ring. See **Fig. 6**, again for the case of $(h,k) = (2,1)$. Such fulleroids always have[b]: 4 triangular rings, $4 \cdot (h^2 + hk + k^2)$ vertices, and $2 \cdot (h^2 + hk + k^2 - 1)$ hexagonal rings.

Next one imagines that each of the triangular faces is contracted to a single point. As a consequence there remain no triangular faces and each of the hexagonal faces adjacent to an original triangular face becomes a pentagon, thereby leaving a fullerene of tetrahedral symmetry. See **Fig.7**. The tetrahedral 3-fold axes pass thru a vertex $i$ where a triple of pentagons meet, and where the net combinatorial curvature[c] (for $i$ & its 3 neighbor sites) is $\pi$, while the combinatorial curvature is $0$ at each of the remaining vertices (which are incident solely to hexagonal rings). The net combinatorial curvature for the polyhedron as a whole then is $4 \cdot \pi$, which matches the net Gaussian curvature of the (smooth) surface on which the fullerene network may be imagined to be embedded. To match combinatorial & Gaussian curvatures locally the Gaussian curvature should then be localized near these same 4 sites (each shared amongst 4 pentagons). And this is neatly accomplished with a lessening of bond-angle stress if each of these 4 sites is modified to be tetrahedrally (~sp$^3$) hybridized, by adding an H atom to each.

Before the placement of H atoms one has a fullerene, which generally are polyhedral structures of trigonal (nominally sp$^2$) sites with all faces either pentagons or hexagons – and it is well-known that then there are exactly 12 pentagons.). As a consequence our fulleroids $TH_4$ are obtained as the tetrahydrogenation of the unique tetrahedral-symmetry fullerenes with 4 sets of 3 incident pentagons. And the mode of strain relief from the near exact match of the Gaussian curvature to combinatorial curvature at the sites near each corner of the tetrahedron seems promising. Thus these $TH_4$ fulleroids seem quite unique, as to curvature matching and stress relief via hybridization & accompanying polyhedralization of the super-tetrahedron.

A further question concerns the stability of the $\pi$-network remaining after the hydrogenation (to give $TH_4$). One simple approach to this is by way of Clar theory [10], especially as adapted [11] to conjugated-carbon nanostructures beyond Clar's classical benzenoids – and in particular to our present fulleroid structures. Generally it is believed that the most stable conjugated-carbon networks are those for which every $\pi$-center is in a Clar sextet (each such sextet being disjoint to every other), and such exceptional structures may be termed [7,12]*Claromatic*. As it turns out, those with $h - k$ being divisible by 3 are[d] precisely the Claromatic species within our present class of fulleroids.

**Tetrahedral Adamanto-Fullerene Units**

A further construction based on suitable tetrahedral symmetry fullerenes with "adamanto" caps is also possible. We start with the construction of the previous section which just before adding 4 H atoms to a considered tetrahedral symmetry fullerene, with triples of pentagonal faces meeting at corners of a tetrahedral structure. Each such fullerene has 4 triples of pentagons otherwise being surrounded by a sea of hexagons. One such triple of pentagons is shown in the first part of **Fig.8**, whereupon we delete the central vertex (shared amongst 3 pentagons) along with its 3 neighbor vertices, and finally link up pairs of vertices adjacent to each one of these neighbors, to give the final result in **Fig.8**. This result has a hexagon centered at each corner of our resultant tetrahedral fullerene, where-after we now cap each such hexagon with a H–C((CH$_2$–)$_3$ unit, connecting this by 3 single bonds to alternate sites of the central hexagon. The so-capped fullerenes have isolated pentagons adjacent to the adamanto-capping, and the $h,k$-labelling must have $h + k \geq 2$. The smallest such adamanto-capped fullerene is based on buckminsterfullerene (i.e., C$_{60}$), and the result C$_{60}$[(CH$_2$)$_3$CH]$_4$ is shown in **Fig.9**.

These species also seemingly have a moderately reasonable matching between combinatorial & Gaussian curvature.

**Tetrahedral Supersymmetries givingTetrahedral Units**

Yet another simple organic unit forming tetrahedral units is cubane, taking the1,3,5,7-positions as the corners of the tetrahedron. Linear units intermediating between the cubanes could also be introduced – including an acetylenic linkage, or a *para*-linked benzene ring, or another cubane linked twice thru positions *para* to one another. And these various nominally linear units may be combined with one another – say with acetylenic units on both sides of a para-linked benzene or cubane units (so as to keep these bulkier straight chain units from interfering with one another in the super-polyhedron). This then leads to a sequences of tetrahedral units as one varies the linear units introduced.

And another sort of possibility is to form the tetrahedral unit from dodecahedrane, appropriately choosing 4 of the 20 CH groups to identify the corners of the tetrahedron. Again bridging linear units could be introduced, but perhaps best just the acetylenic bridges to avoid excess stress (because of steric hindrance problems).

And a yet further possibility for a tetrahedral unit is based on $C_{60}$– which being of icosahedral symmetry has 10 3-fold axes, each passing through the centers of opposite pairs of hexagons. There are 5 subsets of 4 axes, each subset generating a tetrahedral group, whence for one such subset, one may identify 4 hexagons at the corners of a large tetrahedron. Imagining each of such 4 hexagons as a (geometric) base one may add adamantocapsjust as in the preceding section.

Yet also beyond adamantine, there are polymantanes whichmanifest tetrahedral symmetry and could conceivably be used as tetrahedral units. Here one may recognize that for each saturated hydrocarbon X suitably embeddable on the diamond lattice one may imagine an analogous polymantane $p(X)$ constructed by imagining each C atom of X to be centered in an adamantane unit, with neighboring C atoms of X giving rise to adamantane units with a common (hexagonal) ring. So if X has tetrahedral symmetry, then so also will $p(X)$, and further will be a potential tetrahedral unit T. Thus in **Fig. \*\*** we show the carbon skeletons for X being tetramethyl-methane along with the corresponding polymantane $p(X)$.

Finally we might recall the adamantanereplacement operation $a$ discussed a couple sections back. Basically choosing any of the multiplicity of tetrahedral units $T \& T'$ identified in the present or earlier sections, one can entertain $a(T, T')$ as a new tetrahedral unit.

**Stress & Strain**

The angles between adjacent interconnecting edges of a dodecahedron (ideally) are 108º, whereas those for tetrahedral symmetries are $\approx 109.5^a$. Thence there is some degree of stress & strain involved in putting together tetrahedral units to make a super-dodecahedron. Recently something like this has been illustrated[13] in the context of carborane icosahedra, where 12 $C_2B_{10}H_{12}$ icosahedra were combined to form a super-icosahedron, deleting 5 H atoms from each carborane to form 5 intercarborane covalent bonds from each carborane to its 5 neighbor. As a consequence, there is a bond-angle mismatch, ideally ~63.2º(the angle at an icosahedron center between the directrixes thru adjacent icosahedron corners)for each carborane, which in the supercarborane is modified to an equilateral-triangle bond angle of 60º. With this modest mismatch of bond angles, stress was noted – and even more extremely whenthe larger super-icosahedral structures were combined (to a super-super-dodecahedron).

A manifestation of the stress from bond-angle strain is found in the patterning of vibrational levels, with a lower group of levels corresponding to the number of "external" vibrational motions of the building blocks as a whole. Given $\#_{bb}$ building blocks, each with 3 translational & 3 rotational modes of motion, the total number of "external" vibrational modes is

$$\#_{ext} = \#_{bb} \cdot (3+3) - (3+3)$$

where the final 3+3 is subtracted off to eliminate the overall translations & rotations of the overall superstructure. In the case of the carboranes, there were $\#_{bb} = 12$ building blocks, so that there were $\#_{ext} = 66$ external vibrational modes in the range from $120\,\text{cm}^{-1}$ to 350 cm$^{-1}$, after which there was a $\approx 100$-cm$^{-1}$ gap to the 615 remaining "internal" vibrational modes. In our current case, we have $\#_{bb} = 20$ adamantanyl building blocks, so that we have $\#_{ext} = 114$ "external" modes, turning out to be in the range from ~90 cm$^{-1}$ to 330 cm$^{-1}$ (at an average spacing of $\approx 2$ cm$^{-1}$), after which there is a $\approx 30$-cm$^{-1}$ gap to the 917 remaining "internal" vibrational modes. Here the gap is smaller than for the carborane case evidently because the present strain is less (with an angle off-set of just $\approx 1.5^\mathbf{o}$, vs. $\approx 3.2^\mathbf{o}$ for our earlier carborane case).

There is a simple scheme to aid in the relief of stress -- namely to introduce acetylenic linkages between the tetrahedral adamantanyl units. That is, each C—C bond between 2 adamantyl building blocks is replaced by an acetylene bridge

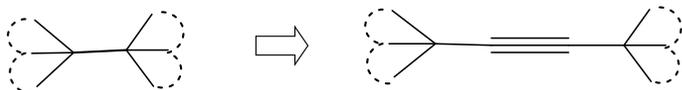

where the dashed lines on the left & right summarily indicate an adamantane cage. Indeed just such a replacement has been suggested previously, in related contexts: in graphene [14], in cubane [15], in different polyhedranes [16,17] and miscellaneous other carbo-cages [18] (including adamantane). Basically the total stress is reduced because the strain is spread out over 3 bonds & 4 bond angles, rather than 1 bond & 2 bond angles, so that with the stress energy quadratic in the strain, the resultant total stress energy should be roughly half as big (from comparing $(\delta/4)^2 + (\delta/4)^2 + (\delta/4)^2 + (\delta/4)^2 = \delta^2/4$ to $(\delta/2)^2 + (\delta/2)^2 = \delta^2/2$, with $\delta$ being the total angle off-set). Inserting a second acetylene into the linkage should drop the stress energy to roughly 1/3 the value with a direct C—C bond between the two adamantanes.

**Conclusion**

The possibility of super-dodecahedra built from a variety of different tetrahedral units has been elucidated. It has been made evident that there is an immense number of possibilities for tetrahedral units, satisfying traditional organic chemical standards of reasonability.

Some brief note might be made that of the various tetrahedral units discussed some of them manifest different tetrahedral symmetries – particularly $\mathcal{T}$ & $\mathcal{T}_d$ (or even $\mathcal{T}_h$). The adamantane based ones as well as most of the others manifest $\mathcal{T}_d$ symmetry. But for our $h,k$-fulleroids it is only those with $k = 0$ or $k = h$ which give $\mathcal{T}_d$. For those fulleroids with $k \neq 0$ or $h$, there then are mirror image forms, and if such $\mathcal{T}$-symmetry units are subjected to the adamantane transformation, then further chiral, or diastereomeric tetrahedral units may be obtained. Our adamanto-capped fulleroids give $\mathcal{T}_d$ symmetry.

Another thing which happens with some of the tetrahedral units $TH_4$ in that not only do they have H atoms outwardly directed at the corners of the tetrahedron, but sometimes they $\mathcal{T}_h$ symmetries, or supersymmetries of $\mathcal{T}_h$ (as octahedral $O_h$ or icosahedral $\mathcal{I}_d$). Then such a unit also has H atoms on the opposite sides of the tetrahedral unit directed in the opposite directions. Such is the case if $TH_4$ is cubane or dodecahedrane, or some of the polymantanes mentioned. Or yet also there is the possibility of a C$_{60}$-based structure with 8 adamantyl cappings (with an additional 4 cappings on the hexagons opposite to those first capped for $C_{60}[(CH_2)_3CH]_4$. Anyway for such $T_h$-symmetry units one could then entertain a filled super-dodecahedron, with an additional suitably sized tetrahedral unit placed inside a superdodecahedron to bond therein with a replacement of the inward directed H atoms.

That there are many tetrahedral hydrocarbons $TH_4$, suggests the possibility of mimicking the great number of saturated hydrocarbons, generally replacing C with T, to generate whole ranges of super-structures, with super-dodecahedrane $(TH)_{20}$ just a typical example. In such an effort a central question should be just which units show a chemistry paralleling that for C, which is to pose questions: Is the T unit internally stable? What is the possibility for functionalization of $TH_4$? If $TH_4$ is readily functionalized, then how do the reactivities parallel those of (tetrahedrally hybridized) carbon? For if these questions are all answered favorably, then there is a good likelihood for the construction of various super-structures, by design, much as for the already rationally developed synthetic organic chemistry.

**Acknowledgements**

Support is acknowledged (*via* grant BD-0894) from the Welch Foundation of Houston, Texas. We acknowledge extensive use of the MIZTLI super computing facility of DGTIC-UNAM. YPO acknowledges a postdoctoral fellowship from the CONACyT research grant 219993. Financial support from PAPIIT-DGAPA-UNAM research grant IG100616 is acknowledged.

**Table 1.** Adamantane-transformed Structures thru Level $n$

| level $n$ | $\alpha_n$ | $\alpha_{n-1}$ | structures |
|---|---|---|---|
| 1 | 1 | 0 | a |
| 2 | 1 | 1 | $a_{\bullet a}$  $a_{a\bullet}$ |
| 2 | 2 | 1 | $a_{aa}$ |
| 3 | 1 | 1 | $a(\bullet, a_{\bullet a})$  $a(\bullet, a_{a\bullet})$  $a(a_{\bullet a}, \bullet)$  $a(a_{a\bullet}, \bullet)$ |
| 3 | 1 | 2 | $a(a, a_{\bullet a})$  $a(a, a_{a\bullet})$  $a(a_{\bullet a}, a)$  $a(a_{a\bullet}, a)$ |
| 3 | 2 | 1 | $a(\bullet, a_{aa})$  $a(a_{aa}, \bullet)$ |
| 3 | 2 | 2 | $a(a_{\bullet a}, a_{\bullet a})$  $a(a_{\bullet a}, a_{a\bullet})$  $a(a_{a\bullet}, a_{\bullet a})$  $a(a_{a\bullet}, a_{a\bullet})$  $a(a, a_{aa})$  $a(a_{aa}, a)$ |
| 3 | 2 | 3 | $a(a_{\bullet a}, a_{aa})$  $a(a_{a\bullet}, a_{aa})$  $a(a_{aa}, a_{\bullet a})$  $a(a_{aa}, a_{a\bullet})$ |
| 3 | 2 | 4 | $a(a_{aa}, a_{aa})$ |

**Figures**

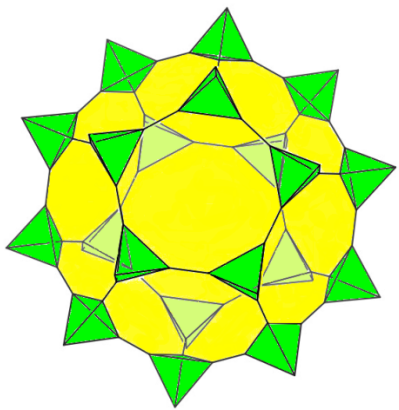

**Fig. 1** –The general construction using 20 tetrahedral units to form a super-dodecahedron. Just tetrahedra on the front side are seen, though the lines in the yellow dodecahedron faces are those on the back side.

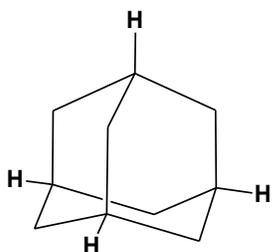

**Fig. 2** –Adamantane, with the 4 H atoms at the tertiary carbons shown in bold-face, and the H atoms on secondary carbons omitted.

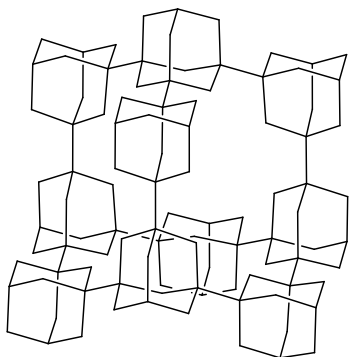

**Fig.3**–Super-adamantane.

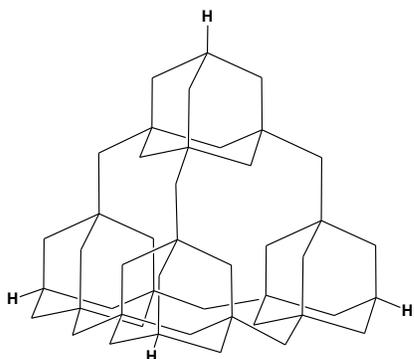

**Fig.4**– Two tetrahedral intermediates between the structures of **Fig.2** & **3**, again with the H atoms at the new tertiary carbons shown in bold-face.

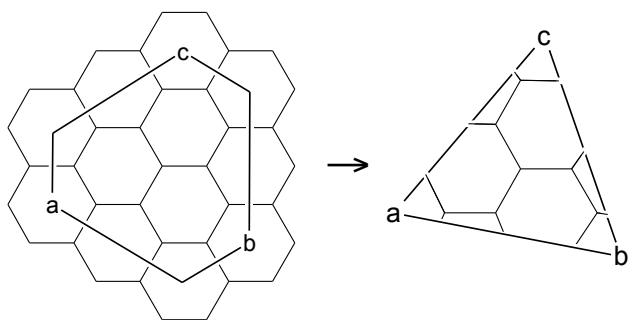

**Fig.5** – The locationing of 3 centers $a, b, c$ for the case of $(h,k) = (2,1)$, then cutting out the equilateral triangle.

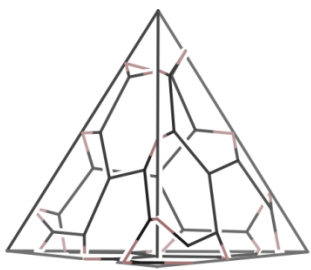

**Fig.6** – An $(h,k) = (2,1)$ tetrahedron.

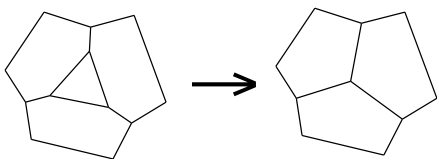

**Fig.7** –The contraction of a triangular face to a single point, and the accompanying formation of 3 pentagons.

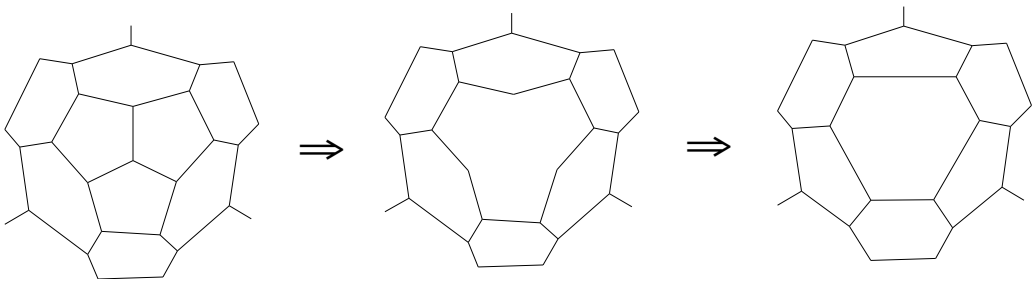

**Fig. 8** –The restructuring around a 3-pentagon vertex.

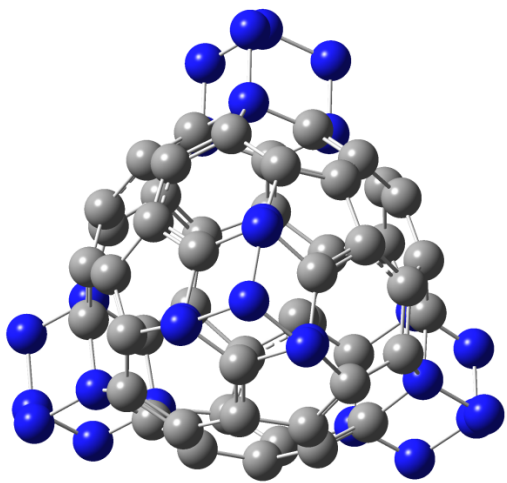

**Fig. 9** –The $C_{60}$-based adamanto-fullerene, with the blue atoms showing the adamanto-caps. The H atoms attached to these adamanto-caps are not shown.

# References

1. Minyaev RM, Popov IA, Koval VV, Boldyrev AI, Minkin VI (2015) Supertetrahedral $B_{80}H_{20}$, $C_{80}H_{20}$, and $Al_{80}H_{20}$ analogs of deocachedrane and their substituted molecules. StructChem 26:223-229

2. Paquette LA, Ternansky RJ, Balogh DW, KentgenG (1983) Total synthesis of dodecahedrane. J Am ChemSoc 105(16):5446-50

3. Semenov SG, Sigolaev FY, Belyakov AV (2013) Adamant-1,3,5-triyl Super Dodecahedranes. StructChem 54:960-962

4. Klein DJ, Bhattacharya D, Panda A, Griffin LL, (2015) Adamantyl Super-structures: Hyper-Adamantane, Hyper-Hyper-Adamantane, & Beyond, to Fractality. Intl J Chem Mo. 6:221-230

5. Huang M, Hsu CH1, Wang J, Mei S, Dong X, Li Y, Li M, Liu H, Zhang W, Aida T, Zhang WB, Yue K, Cheng SZ(2015).Science 348(6233):424-8

6. Kaszynski P, Michl J (1988): [n]Staffanes: A Molecular-Size 'Tinkertoy' Construction Set for Nanotechnology. Preparation of End-functionalized Telomers and a Polymer of [1.1.1]Propellane. J Am ChemSoc 110:5225-5226

7. Zhu CCA (1995)

8. Schleyer PVR, Williams JR Jr, Blanchard KR(1970). J AmerChemSoc 92(8):2377-86

9. Van Vechten D, Liebman JF (1981). Isr JChem 21(2-3):105-10

10. Clar E (1971) The Aromatic Sextet. John Wiley & Sons, NY

11. Klein DJ, Balaban AT (2011) Clarology for Conjugated Carbon Nano-Structures: Molecules, Polymers, Graphene, Defected Graphene, Fractal Benzenoids, Fullerenes, Nano-Tubes, Nano-Cones, Nano-Tori, *etc*. Open J Org Chem 5:27-61 (Suppl 1-M3)

12. Balaban AT, Klein DJ (2009) Claromatic Carbon Nano-Structures. J PhysChem C 113:9123-19134

13. Bhattacharya D, Klein DJ, Oliva JM, Griffin LL, Alcoba, DR, Massaccesi GE (2014) Icosahedral symmetry super-carborane and beyond. ChemPhysLett 616-617:16-19

14. McDiarmid, 1994

15. Manini P, Amrein W, Gramlich V, Diederich F (2002) AngewChem Intl Ed 41(22):4339-4343

16. Jarowski PD, Diederich F, Houk KN (2005). J OrgChem 70(5):1671-1678

17. Bachrach SM, Demoin DW (2006). J Org Chem 71(14):5105-5116

18. Azpiroz JM, Islas R, Moreno D, Fernandez-Herrera MA, Pan S, Chattaraj PK, Martinez-Guajardo G, Ugalde JM, MerinoG (2014) Carbo-cages: A Computational Study. J Org Chem 79(12):5463-5470
10

**Appendix**

---

[a] To obtain the number $\#_{@n}$ of structures with iteration thru (exactly) to level $n$, we also introduce the number $\#_{<n}$ of structures with iteration not going thru to level $n$ (which is to say thru only to some level $m$ with $m < n$). Then to obtain the count $\#_{@n+1}$ for level-$n+1$ structures, we are interested in counting up the transformations $a(x,y)$ where at least 1 of the structures $x$ or $y$ is a level-$n$ structure: the number of possibilities where both $x$ & $y$ are level-$n$ structures is simply $\#_{@n} \cdot \#_{@n}$, while the number where $x$ is of level $n$ & $y$ is less than $n$ is $\#_{@n} \cdot \#_{<n}$, and finally the number where $x$ is less than $n$ & $y$ is at level $n$ is $\#_{<n} \cdot \#_{@n}$. That is, we have $\#_{@n+1} = 2\#_{@n}\#_{<n} + \#_{@n}^2$. But clearly we also have $\#_{<n+1} = \#_{@n} + \#_{<n}$. Starting from $\#_{@1} = \#_{<1} = 1$, we iterate this pair of recursions to find: $\#_{@2} = 3$, $\#_{@3} = 21$, $\#_{@4} = 651$, $\#_{@5} = 457653$, etc. Asymptotically $\#_{@n} \sim \kappa^{(2^n)}$, for some $\kappa > 1$, and further $\#_{<n} \sim (\sqrt{\kappa})^{(2^n)}$.

[b] The construction of triangular faces follows that performed elsewhere* to make icosahedral-symmetry fullerenes – with such triangular patches from the graphene lattice being placed on the faces of a large icosahedron. There the number $n$ of vertices of a patch is identified as $h^2 + hk + k^2$. It is clear that there are 4 triangles in the present construction. And further it is seen that every vertex of the fulleroid is in 3 faces, while each $m$-membered ring adjoins to $m$ vertices, so that the number of incidences between vertices and faces is given by both $3n$ & $3f_3 + 6f_6$, with $f_m$ the number of $m$-sided faces. Setting $f_3 = 4$, we can solve for $f_6$ to obtain the claimed result.

[c] The combinatorial curvature $\kappa_i$ at a vertex $i$ in a graphical structure (cellularly) embedded on a surface $S$ may be defined* as $\kappa_i \equiv 2\pi - \sum_{\alpha}^{@i}(\pi - \frac{2\pi}{|\alpha|})$ where the sum is over the different faces meeting at $i$, and $|\alpha|$ is the number of sides to face $\alpha$. Standardly the linear curvature $\lambda(x)$ of a smooth curve $f(x) = y$ at $x$ is $\pm$ the inverse of the radius of the circle tangent (at $x$) to $f(x)$ with the circle defined such that its second derivative matches that of $f(x)$ – and the $\pm$ sign is chosen to indicate on which side of the tangent line the circle center resides. Also standardly (in differential geometry) Gaussian curvature at a point $p$ on a smooth surface $S$ entails linear curvatures $\lambda$ & $\lambda'$ for two curves which are the intersections of $S$ with two planes containing the normal to $S$ at $p$ -- the two planes being chosen to be orthogonal to one another. Then the Gaussian curvature $\gamma_p$ at a point $p$ on $S$ is the product of the extreme values for such $\lambda$ & $\lambda'$. Another way to define $\gamma_p$ takes the normal direction to be a (local) $z$-axis at $p$, and locally around $p$ views $S$ as a function of $x$ & $y$, to identify $\lambda$ & $\lambda'$ as the eigenvalues of the $2 \times 2$ matrix $\begin{pmatrix} \partial^2 S/\partial x^2 & \partial^2 S/\partial x \partial y \\ \partial^2 S/\partial y \partial x & \partial^2 S/\partial y^2 \end{pmatrix}$ with the derivatives evaluated at $p$. A classical result of differential geometry, the Gauss-Bonnet theorem, states that the net Gaussian curvature integrated over an arbitrary smooth closed surface topologically equivalent to a sphere is $= 4\pi$. Summation of the combinatorial curvature over a whole polyhedral graphic structure $G$ also gives* $4\pi$, as a consequence of

Euler's theorem that: if $V$, $E$, & $F$ are the respective numbers of vertices, edges, & faces, then $V - E + F = 2$. This gives a rigorous mathematical consequence: global (or net) combinatorial & Gaussian curvatures match exactly. But in addition is seems that local matching gives relief of local stresses (& strains) – so that for the present example most of the Gaussian curvature should be near of the apices of the tetrahedral unit.

[d]To see this one first considers graphene (from which the $h,k$-nets for our fulleroids are developed). Bulk graphene indeed is Claromatic: a fully sextet resonant Clar structure for graphene is obtained with each non-sextet ring surrounded at every other ring by Clar sextet rings. Now taking ring-center to adjacent ring-center steps one may walk in straight lines, except possibly for a single turn of 60° from going straight ahead. Given a fully sextet-resonant Clar structure, it may be shown that [7] every such walk of straight line segments of $h > 0$ & $k \geq 0$ steps, with the walk beginning & ending in a sextet-resonant ring, are exactly those for which $h - k$ is divisible by 3. With the choice of triangular sections of $h,k$-nets used in our construction of our $h,k$-fulleroid, following thru the construction reveals that the remnant $\pi$-network is then fully sextet resonant.